\documentclass[12pt]{article}
\usepackage{amsmath,amssymb,graphicx} 
\def\be{\begin{equation}}
\def\ee{\end{equation}}
\def\bib{\bibitem}
\def\ql{\lq\lq}
\def\qr{\rq\rq}

\def\ef{\eqref}\def\nn{\nonumber}
\def\bq{\begin{quote}}
\def\eq{\end{quote}}
\def\pt{\partial}
\def\nc{\newcommand}
\def\half{{\tfrac{1}{2}}}
\def\a{\alpha}
\def\b{\beta}
\def\G{\Gamma}
\def\k{\kappa}
\def\l{\lambda}
\def\m{\mu}
\def\n{\nu}
\nc{\gamu}[3]{{\G^{#1}}_{{#2}{#3}}}
\begin{document}
\begin{center}
\Large{\bf The Cosmological Constant}\\
\large{Alex Harvey}$^{a)}$ \\
\vspace{.5cm}
\vspace{.15cm}
\normalsize
Visiting Scholar \\
New York University \\
New York, NY 10003 \\
\vspace{.8cm}
\end{center} 
\vspace{.8cm}
\begin{abstract}
Contrary to popular mythology, Einstein did {\em not\/} invent the cosmological constant just in order  construct his model universe.  He discussed it earlier in \ql The Foundations of General Relativity\qr in connection with the proper structure of the source-free field equations.  There he dismissed it as arbitrary and unnecessary.  It was later that he found its inclusion to be essential to the construction of his model.
\end{abstract}
Contrary to virtually universal belief, the cosmological constant was {\em not\/} invented by Einstein \cite{ein1} for use in  his {\it cosmological Considerations Concerning General Relativity\/} \cite{ein2}.  The fact is, he discussed it explicitly about a year earlier in Sect. (14) of {\it The Foundations of General Relativity\/} \cite{ein3} where he introduced the source-free field equations. 
\begin{subequations}\begin{align}     
      \frac{\pt {\G^\a}_{\m\n}}{\pt x_\a} + \gamu{\a}{\m}{\b}\gamu{\b}{\n}{\a}&= 0  \\  
              \sqrt{-g} &= 1 \,.
\end{align}\end{subequations}
(The second equation is a \ql coordinate condition\qr which permits writing the field equations elegantly if severely truncated.  In this form, the field equations have never been applied to any problem.)  There he comments, 
\bq \ql It must be pointed out that there is only a minimum of arbitrariness in the choice of these equations.  For besides $G_{\m\n}$ there is no tensor of second rank which is formed from the $g_{\m\n}$ and its derivatives, contains no derivatives higher than the second and is linear \cite{ein4} in these derivatives.\qr
\eq 
Then he adds a footnote, 
\bq\ql Properly speaking this is true only of the tensor
\be\label{add}
                   G_{\m\n} + \l g_{\m\n} g^{\a\b}G_{\a\b}
\ee
where $\l$ is a {\em constant\/} [emphasis added].  If, however we set this tensor $=0$, we come back again to the equation $G_{\m\n}=0$.\qr 
\eq  

The term in $\l$ is is not the optimal choice.  In modern notation, $G_{\m\n}$ is the Ricci tensor so that Exp. \ef{add} would be
\be
                R_{\m\n} +\l g_{\m\n} R \,.
\ee
This would result in the source-free equations
\be
                                  R_{\m\n} = \frac{1}{4}g_{\m\n}R \,.
\ee
   
The more desirable expression is the simpler $\l g_{\m\n}$.  In fact, the most general tensor satisfying the conditions mentioned earlier and which has a vanishing divergence is \cite{love} is (the now standard)
\be\label{love}
                  R_{\m\n} -\half g_{\m\n}R +\l g_{\m\n} \,.
\ee
In the source-free case this reduces readily to
\be
                R_{\m\n}  = \frac{\l}{4} g_{\m\n}   \,.
\ee
These have been studied by Petrov \cite{petrov} who termed the solutions, \ql Einstein Spaces\qr.
 
Einstein sorted this out in the {\it Cosmological Considerations ...\/}.  In Eq. (13) therein he included the cosmological constant
\be\label{a}
                   R_{\m\n} - \l g_{\m\n} = -\k(T_{\m\n} -\half g_{\m\n}T) 
\ee
which is readily converted to
\be\label{b}
        R_{\m\n} - \half g_{\m\n}R + \l g_{\m\n}\ = -\k T_{\m\n} \nn \,. \\
\ee
${}^{a)}$Prof. Emeritus, Queens College, City University of New York; ah30@nyu.edu.\\

\end{document}